\documentclass[12pt, preprint]{aastex}

\usepackage{natbib}
\newcommand{\Msolar}{M_\odot}
\begin{document}

\title{DR21 Main: A Collapsing Cloud \footnote{Dissertation submitted to the Department of Astronomy and Astrophysics, University of Chicago, in partial fulfillment of the requirements for the Ph.D. degree.}}
\author{Larry Kirby}
\affil{Department of Astronomy and Astrophysics, Enrico Fermi Institute}
\affil{The University of Chicago, Chicago, IL 60637}
\email{lkirby@oddjob.uchicago.edu}

\begin{abstract}
The molecular cloud, DR21 Main, is an example of a large-scale gravitational collapse about an axis near the plane of the sky where the collapse is free of major disturbances due to rotation or other effects.  
Using flux maps, polarimetric maps, and measurements of the field inclination by comparing the line widths of ion and neutral species, we estimate the temperature, mass, magnetic field, and the turbulent kinetic, mean magnetic, and gravitational potential energies, and present a 3D model of the cloud and magnetic field.
\end{abstract}

\keywords{Molecular Clouds, Submillimeter, Polarization, Star Formation}
\maketitle
\section{Introduction}
Partially ionized material in a molecular cloud will respond to gravitational forces by moving, primarily, along magnetic field lines.  If one assumes that the field is not significantly tangled by turbulence, one can expect a concentration of material in an oblate core. This model has been verified by the analysis of polarimetric and photometric data \citep{Dotson} by \citet{tassis} showing that the field is directed along the short axes of the cores. The neutral component of the material will move slowly across field lines toward the center of gravity thus increasing the gravitational field. Where a pinch in the field is indicated by a polarization map, one can infer that the gravitational field in the core is competing with the support provided by the field. Within a critical radius, one expects the cloud to collapse.

Examples of pinched fields have been found in Orion \citep{Schleuning} and in NGC1333 \citep{Girart}. In the case of DR21 Main (DR21M), the polarization map extends to radii such that the field straightens out as the gravitational field diminishes leaving an hourglass field configuration. In this cloud one has an opportunity to estimate the mass within the collapsing region.
We use 350 $\rm \mu m$ polarimetry to measure the field as projected on the sky (\textsection \ref{polarization}); line width measurements to estimate the inclination of the field to the line of sight (\textsection \ref{inclination}); photometric maps to estimate the mass distribution (\textsection \ref{mass}); and the angular dispersion of the field vectors to estimate the field strength (\textsection \ref{bfield}).

\section{DR21}
\label{DR21}
\begin{figure}
\epsscale{0.75}
\plotone{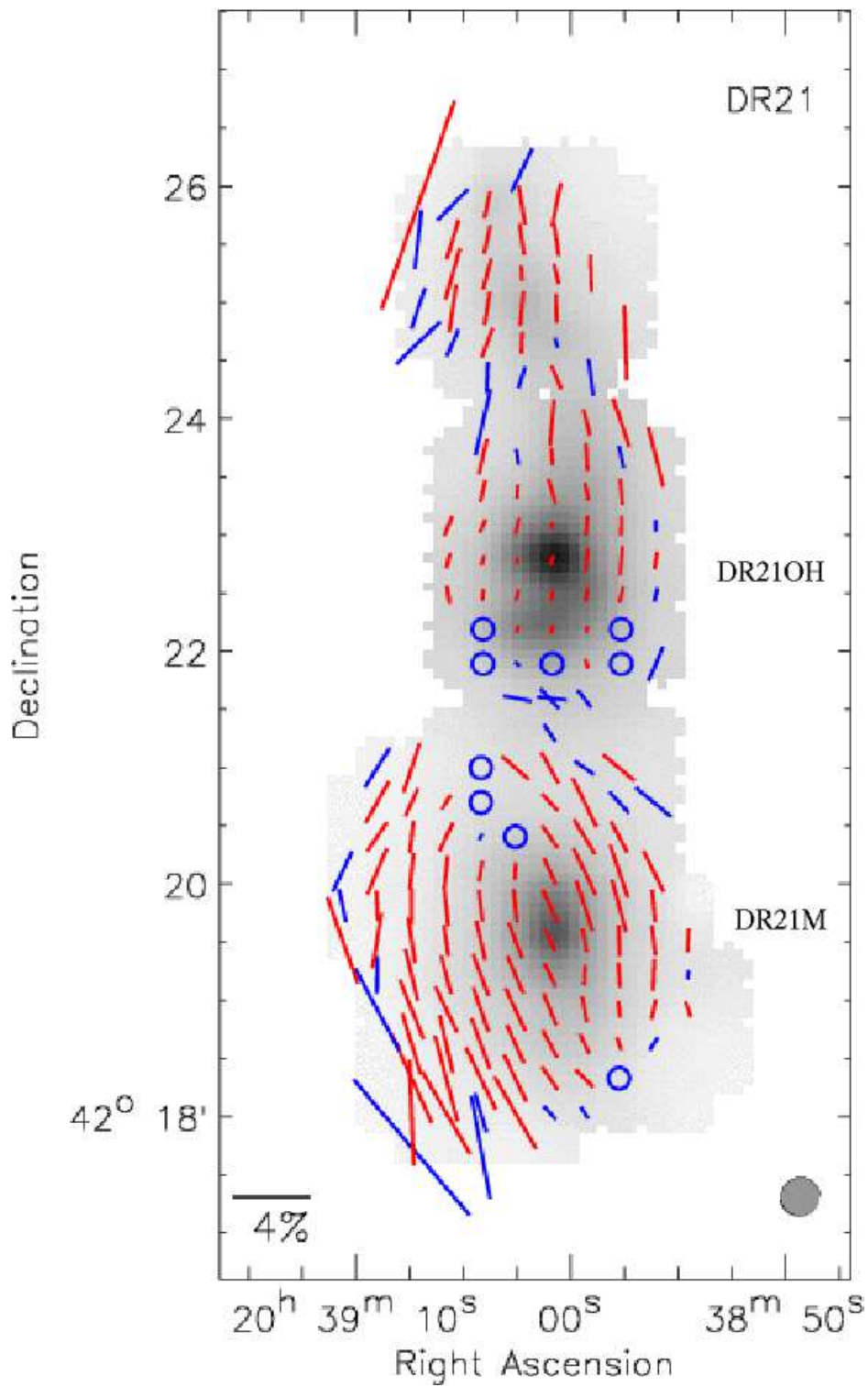}
\caption[DR21 Cloud]{Hertz image \citep{Dotson} of the star forming region DR21.  Red vectors (E-vectors) are for measurements better than 3$\sigma$ polarization and blue vectors are for measurements of 2$\sigma$ to 3$\sigma$.  Open circles are locations with a 2$\sigma$ upper limit of 1\%
\label{Hertz view of Cygnux region}}
\end{figure}

\begin{figure}
\plotone{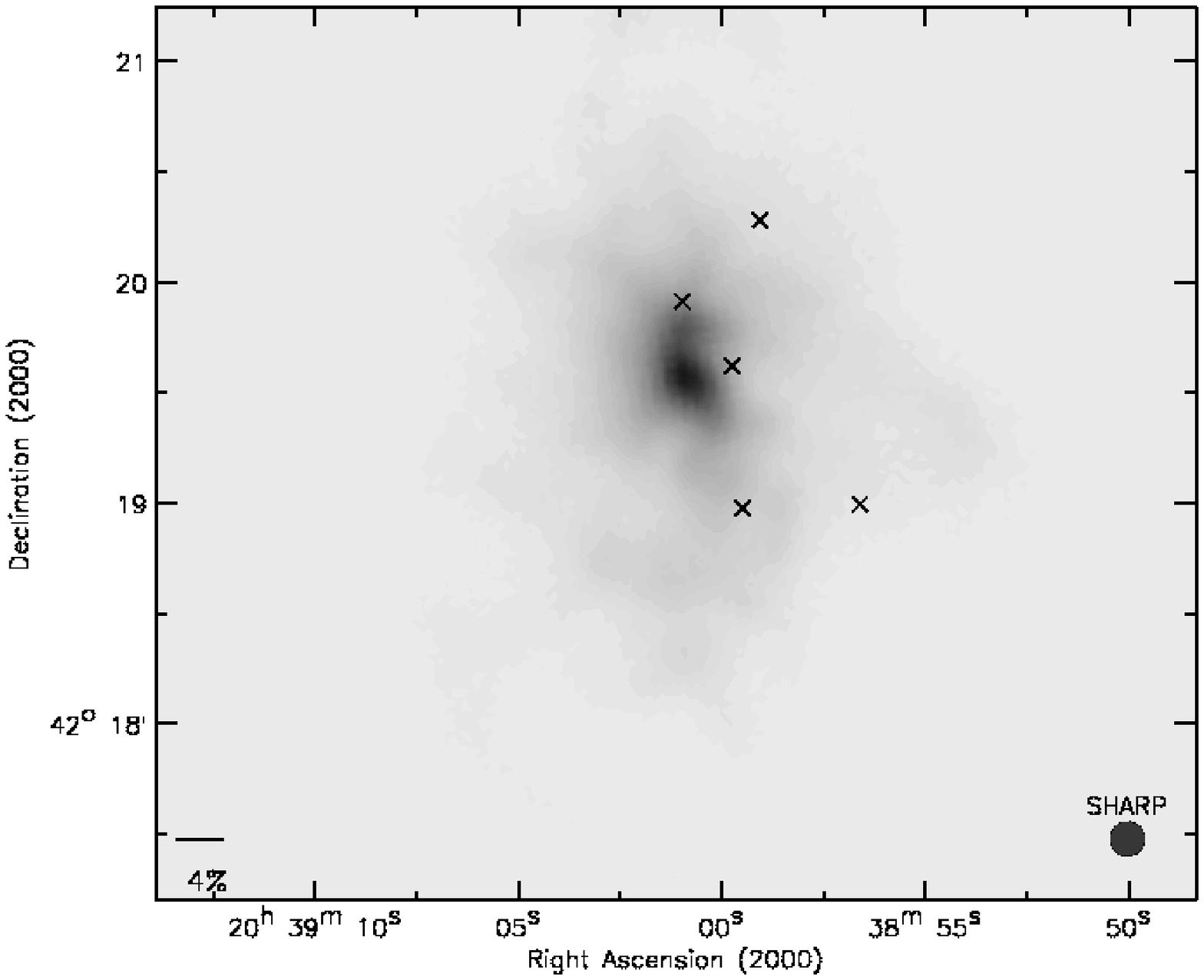}
\caption[DR21 NIR sources]{SHARCII 350 $\mu m$ flux map with x's marking locations of Spitzer NIR sources \citep{kumar07, Davis07}
\label{NIR}}
\end{figure}

The giant star forming complex, DR21 (Figure \ref{Hertz view of Cygnux region}) is located in the Cygnus constellation $\sim $ 3kpc from Earth \citep{Campbell82}.  The southernmost component, DR21M, has a mass of $\rm \sim 20,000 \  \Msolar$ \citep{Richardson}.  It contains one of the most energetic star formation outflows detected \citep{Garden91, Garden92}.  \citet{Garden91} measured the mass of the outflow to be $\rm 3000 \ \Msolar$. Observers using Spitzer found 5 near-infrared (NIR) sources (see Figure \ref{NIR}) inside DR21M, presumed to be protostars \citep{kumar07, Davis07} indicating that regions within the cloud are undergoing gravitational collapse.  \citet{Glenn} at 1.3 mm and \citet{Minchin} at 800 $\rm \mu m$ have previously measured the polarization of DR21 finding uniform position angles $\rm \sim 25 \arcdeg$ East of North (i.e. B field $\rm \approx$ 65 \arcdeg West of North).  The galactic plane is oriented at 42 \arcdeg East of North.  \citet{Roberts} measured a line-of-sight magnetic field of $\sim 400 \mu G$ using HI Zeeman measurements around the dust emission peak and star formation regions.  The regions observed by \citet{Roberts} with $\rm 3.5\sigma$ detections are outlined in green in Figure \ref{sharpandhertz-fits}.

\section{Observations}
\label{observations} 
\subsection{Polarization}
\label{polarization}
 Observations, made at the Caltech Submillimeter Observatory (CSO) with the polarimeter Hertz \citep{Schleuning97, Dowell}, provided 350 $\rm \mu m$ polarization maps of the region.  Hertz was later replaced by SHARP \citep{Li07}, a module mounted ahead of the photometer SHARC-II \citep{sharc} to permit polarization measurements.

SHARP has a wire cross-grid to split the incoming signal into orthogonal components of polarization which are then detected on two 144 pixel subsets of the SHARC-II detector array.   Standard infrared techniques of chopping the secondary mirror to remove sky emission and positional offsets to remove gradients in sky emission were used for all the polarimetry observations \citep{primer}.  The chopping direction was azimuthal with a throw of 300\arcsec.  To avoid chopping into the adjacent cloud, DR21OH, all observations were made when the chop throw was $\rm >$ 30\arcdeg from north-south on the sky  A half-wave plate was rotated to four different angles separated by 22.5\arcdeg.  Each observation cycle consisted of a set of observations at the 4 half-wave plate angles.  The signals from the two 144 pixel arrays were then combined to give the polarization and photometric signals as described by \cite{Kirby05}.  The polarization and flux signals are given by
\begin{eqnarray}
\mathrm{Polarization\  Signal} = H - V\\
\mathrm{Flux\ Signal} = H + V
\end{eqnarray}
where $H$ and $V$ are the horizontal and vertical components of the signal.  The Stokes parameter, $Q$, was calculated from the difference of the polarization signals of the 1st and 3rd angle in the cycle, while the parameter $U$ was calculated from the difference of the polarization signals of the 2nd and 4th angle.  The source flux, $I$,  was determined by the average of the flux signals from the 4 half-wave plate positions.  The $Q$, $U$, and $I$ from each cycle were then combined after correcting for changes in attenuation and sky noise.  Due to the lack of an instrument rotator, this combination had to be done a new way.  The corrected $Q$, $U$, and $I$ from all of the cycles were combined to form an irregularly sampled map.   This map was then smoothed to a resolution of $\rm \sim 10 \arcsec$ (for more detail see \cite{martinjohn}.  

The observations were made in 7 fields spaced by 45\arcsec.  Each field was observed with a 4-point dither on corners of a square 10\arcsec \ wide in Right Ascension and Declination.  A few additional observations were taken to fill in areas of low signal to noise in the above 7 fields.  The data were corrected for instrument polarizations of $\rm \sim 0.3-0.5 \%$ (\citet{Li07}, Vaillancourt et al, in preparation)

\begin{figure}
\plotone{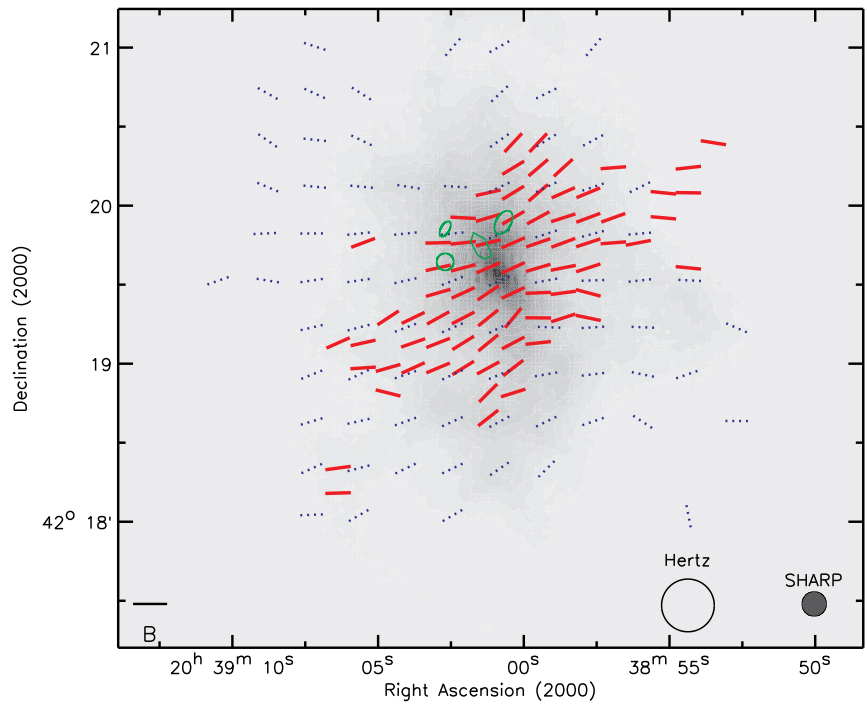}
\caption[SHARP and Hertz Polarization] {350 $\rm \mu m$ SHARP(red) and Hertz(blue dotted) polarization of DR21M.  B vectors are all set to the same arbitrary length in the magnetic field direction.  The grayscale background comes from a SHARC-II flux map.  Green contours show the locations larger than a SHARC-II pixel with 3.5$\sigma$ Zeeman results from \citet{Roberts} 
\label{sharpandhertz-fits}}
\end{figure}

The SHARP observations were taken on 2006 July 15-19 during intermediate weather ($\rm \tau_{225GHz} \approx 0.07-0.08$ corresponding to 1.7-2.0 mm zenith water vapor).  The SHARP polarization data set shows 1270 sky positions with a signal to noise $\rm \geq 3$.  The beam size of SHARP is $\rm \sim {4} \times$ the pixel size.  Limiting the sample so there is no beam overlap leaves 78 measurements.  The median value of the polarization is 2.4\% ranging between 0.37\% and 13\%.  Half of the points have a signal to noise better than 4.7.  At this value, the accuracy of the position angle of the polarization vectors is $\rm \sim 6.5 \arcdeg$.  The direction of polarization shows a pinched or hourglass shape (see Figure \ref{sharpandhertz-fits}).   The axis of symmetry of the hourglass is inclined $\rm \sim$ 15\arcdeg \ East of North (determined by model see \ref{3d}.

The results tabulated by \citet{Dotson} were obtained with the polarimeter, Hertz (Figure \ref{sharpandhertz-fits} blue dotted vectors).  The Hertz measurements were taken on three observing runs, 1997 April 18-27, 1997 September 18-26, and 2001 April 10-16. 

\subsection{Heterodyne Measurements}
The polarimetry measurements from SHARP and Hertz are supplemented with 345 GHz heterodyne measurements of ion and neutral molecular line widths from the CSO.  The heterodyne measurements were taken on 2006 July 12-13  ($\tau_{225GHz} \approx 0.1$).  The inclination of the field at eleven points within the pinched region was determined from measurements of line widths of neutral and ionized molecules of comparable mass.  The heterodyne measurements of the J$\rm \rightarrow$4-3 transition in HCN and HCO+ were made using only on/off position differencing \citep{Houde00,Houde002}.  The data were reduced using the CLASS program as part of the GILDAS package of programs (http://www.iram.fr/IRAMFR/GILDAS).
  Each of the lines was smoothed to half of the natural resolution using the CLASS routine SMOOTH to improve the signal to noise.  One of the positions was removed due to low signal to noise.  Another position was unusable because bandwidth of the instrument was insufficient to image the entire line. 

\begin{figure}
\plotone{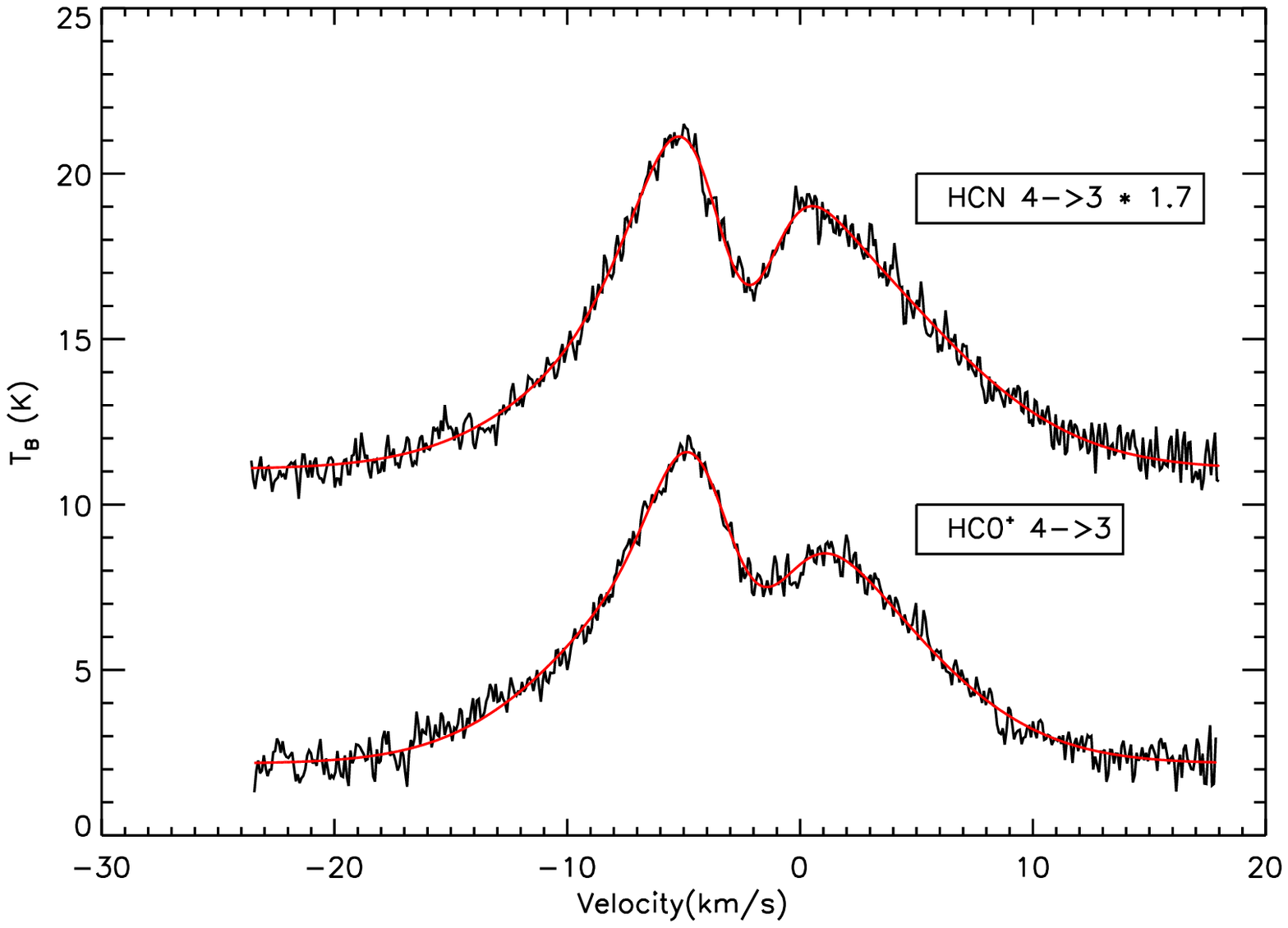}
\caption[Line Profiles of DR21M Center] {Line profile (black) and best fit multi-Gaussian model (red) of the J$\rm \rightarrow$4-3 transition of HCN (top) and J$\rm \rightarrow$4-3 transition of HCO$\rm^+$ (bottom) at the center of DR21M (20$^h$37$^m$14.1$^s$, 42\arcdeg8\arcmin53\arcsec).
\label{linewidths}}
\end{figure}

The spectrum at each position was then fit by a multi-Gaussian model (Figure \ref{linewidths}) to determine the line widths (more properly the standard deviation $\rm \sigma_v$).  Multiple Gaussians were used to account for the self-absorption seen in most of the positions, and for one-sided outflows, which can cause an artificial reduction in line widths if not accounted for properly.  The effects from broad outflows were removed from two locations.  The fits were done using the XGaussfit program from the FUSE package of software for IDL ( http://fuse.pha.jhu.edu/analysis/fuse\_idl\_tools.html).  Results of the fits are given in Table \ref{iatab}.

\begin{deluxetable}{cccccc}


\tablecaption{Line Width Data and Inclination Angle with the Line of Sight for DR21 
\label{iatab}}
\tablehead{\colhead{$\Delta$RA\tablenotemark{a}} &
\colhead{$\Delta$Dec\tablenotemark{a}} & \colhead{$HCN\ \sigma_v (km/s)$} &
\colhead{$HCO^+\ \sigma_v (km/s)$} & \colhead{$P(Hertz 350 \mu m)$} & \colhead{$\alpha (deg)$}
}
\startdata
0  &   0& 6.4& 6.2& 1.2 &39.3\\    
20  &  0 &4.9 &4.8& 1.95 &28.3\\   
-20 &  0 &5.8 &5.7& 1.24 &48.7\\ 
0  & -20 &4.8 &4.4& 1.36 &45.2\\
0  &  20 &3.5 &3.2& 1.84 &43.5\\
-20& -20 &5.4 &4.3& 1.09 &61.6\\
-20&  20 &2.0 &1.6& 2.72 &62.4\\
20 &  20 &2.4 &1.7& 1.74 &68.4\\
20 & -40 &2.5 &1.3& 2.46&77.1\\
\enddata

\tablenotetext{a}{Position Offsets in arcseconds from $20^h37^m14.1^s$, $42\arcdeg8\arcmin53\arcsec$ (2000).}

\end{deluxetable}

\section{Results} 
\label{cha:res} 
\subsection{Inclination Angle} 
\label{inclination}
\begin{figure}
\plotone{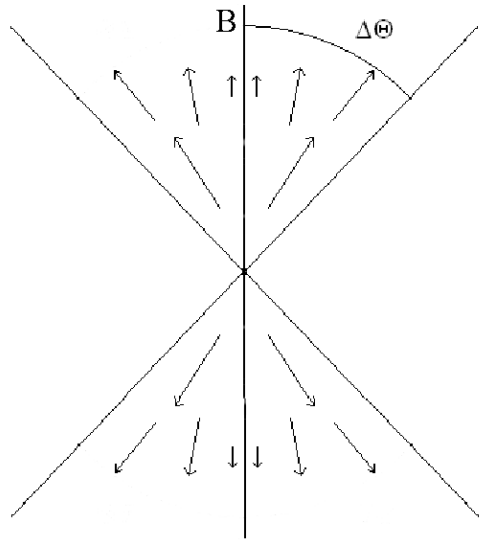}
\caption[Neutral material flow]{Illustration of flow of neutral material.  All flows (velocity vectors) are contained within a cone of width $\rm \Delta\theta$ centered on the symmetry axis defined by the magnetic field (see Figure 2 of \cite{Houde02}.  Length of vectors is arbitrary as magnitude and direction are assumed independent.
\label{nfimage}}
\end{figure}

\cite{Houde00,Houde002} investigated the effect of a magnetic field on the velocities and ions in a partially ionized flow (not necessarily an outflow or jet but any situation where local mean velocity is not zero).  For a significantly strong field, the ions were forced into gyromagnetic rotation about the magnetic field instead of following the flow.  This rotation led to a reduction in line width and suppression of high velocity wings.  The line widths were derived for both the neutrals and ions.
The ratio of the linewidths depends only on the orientation of the neutral flow(s) and the inclination angle of the magnetic field with the line of sight \citep{Houde02,Houdeetall04}.  Assuming that the neutral component of the material is $\rm H_2$ (mean molecular mass of 2.3), the square of the ratio of the ion to neutral line width is given in terms of the angle with the line of sight, $\rm \alpha$, by
\begin{equation} 
\frac{\sigma^2_{l,i}}{\sigma^2_{l,n}} \approx \frac{e\ \cos{\alpha}^2+f(0.16\cos{\alpha}^2+0.84\sin{\alpha}^2/2)(m_i/\mu_i-1)^{-1}}{e\ \cos{\alpha}^2+f\sin{\alpha}^2/2} 
\label{crap}
\end{equation} 
where
\begin{eqnarray}
e=\frac{1-\cos^3{\Delta\theta}}{6},\\
f=\frac{2-3\cos{\Delta\theta}+\cos^3{\Delta\theta}}{6},
\label{crap2}
\end{eqnarray}
and $m_i$ and $\mu_i$ are the ion mass and the reduced mass respectively.  The neutral flows are modeled as symmetrical and contained within a cone of width $\Delta\theta$ centered on magnetic field direction (symmetry axis, see Figure \ref{nfimage}).  All values of $\alpha$ must satisfy the condition $P/P_{max} \leq \sin{\alpha}^2$ where $P_{max}$ is the maximum polarization seen by Hertz (10 \% \citet{Houdeetall04}).  

Although the cone half width, $\Delta\theta$, can, in principle, be determined by finding the curve that best fits the above condition on $P/P_{max}$, the data for DR21 do not provide significant constraints on this quantity.  The values of $\alpha$ presented in Table \ref{iatab} are computed assuming $\Delta\theta = 45\arcdeg$.  The inclination with the line of sight decreases toward the center of DR21 as would be expected for a magnetic field in the center of a gravitationally contracting region with an axis near the plane of the sky.  Inclination angles for points located at the edge of the cloud, where the gravitational pinch has had only a small effect, should approximate the inclination of the cloud's  mean B field.  These points for DR21 give an estimate of $\rm \sim 70\arcdeg$ inclined to the line of sight or $\rm \sim$ 20\arcdeg to the plane of the sky.  

Choosing the wrong value of $\Delta\theta$ leads to a systematic error in the calculation of each angle, but the difference between angles is not significantly affected.  The overall trend of decreasing angle toward the center of the cloud remains.  If, instead, we had assumed $\Delta\theta = 90\arcdeg$, the values shown in the table would decrease by $\sim 15\arcdeg$ but the overall trend toward decreasing inclinations toward the center would not change.

\begin{figure}
\plotone{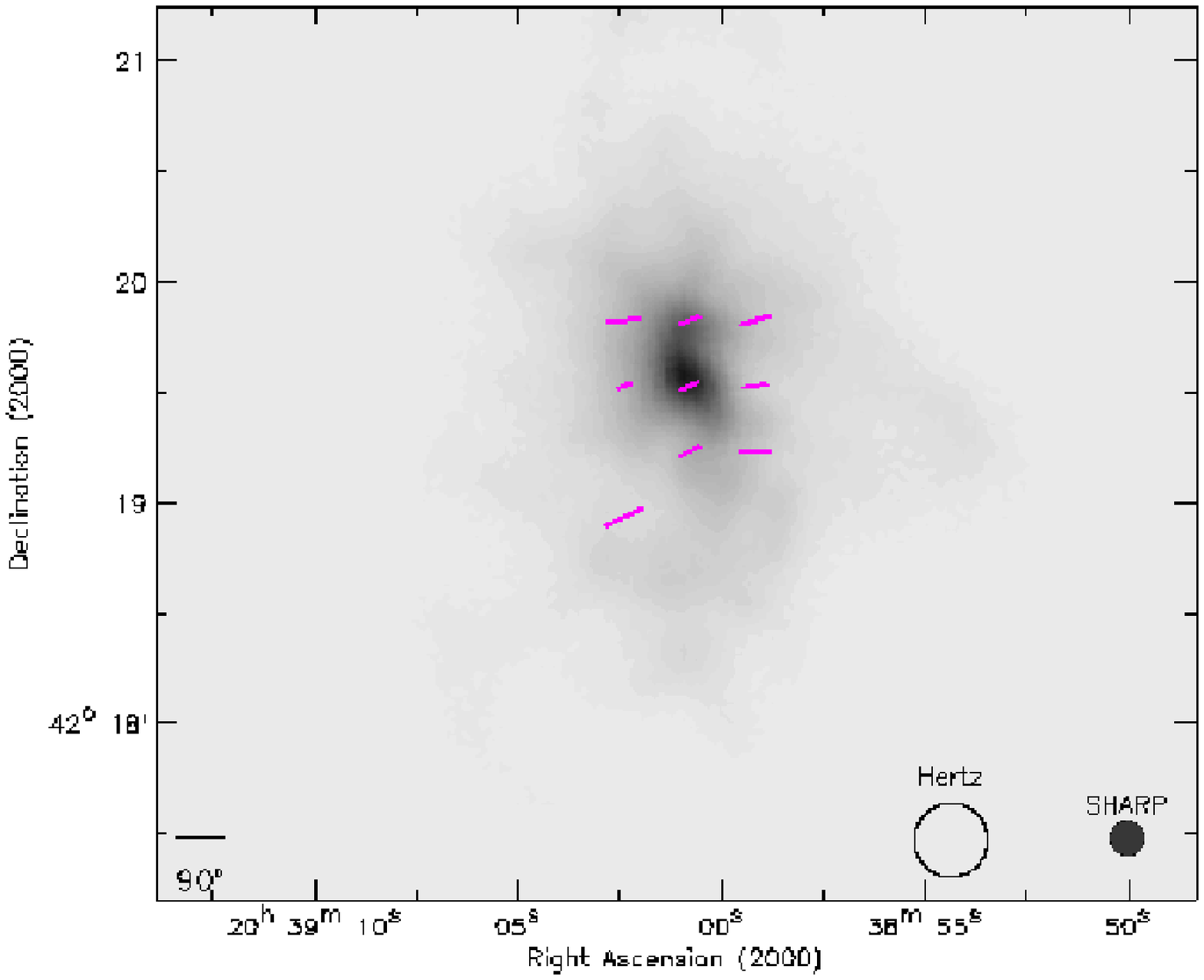}
\caption[3D orientation of the magnetic field] {Polarization vectors with known 3-dimensional spatial orientation.  The angle of the vectors is drawn at the angle of the magnetic field in the plane of the sky.  The length of the vector is proportional to the angle, $\rm \alpha$, of the magnetic field with respect to the line of sight.  The grayscale background is from SHARC-II.
\label{incangle}}
\end{figure}

\subsection{Temperature and Optical Depth}
\label{tempandtau}
\begin{figure}
\plotone{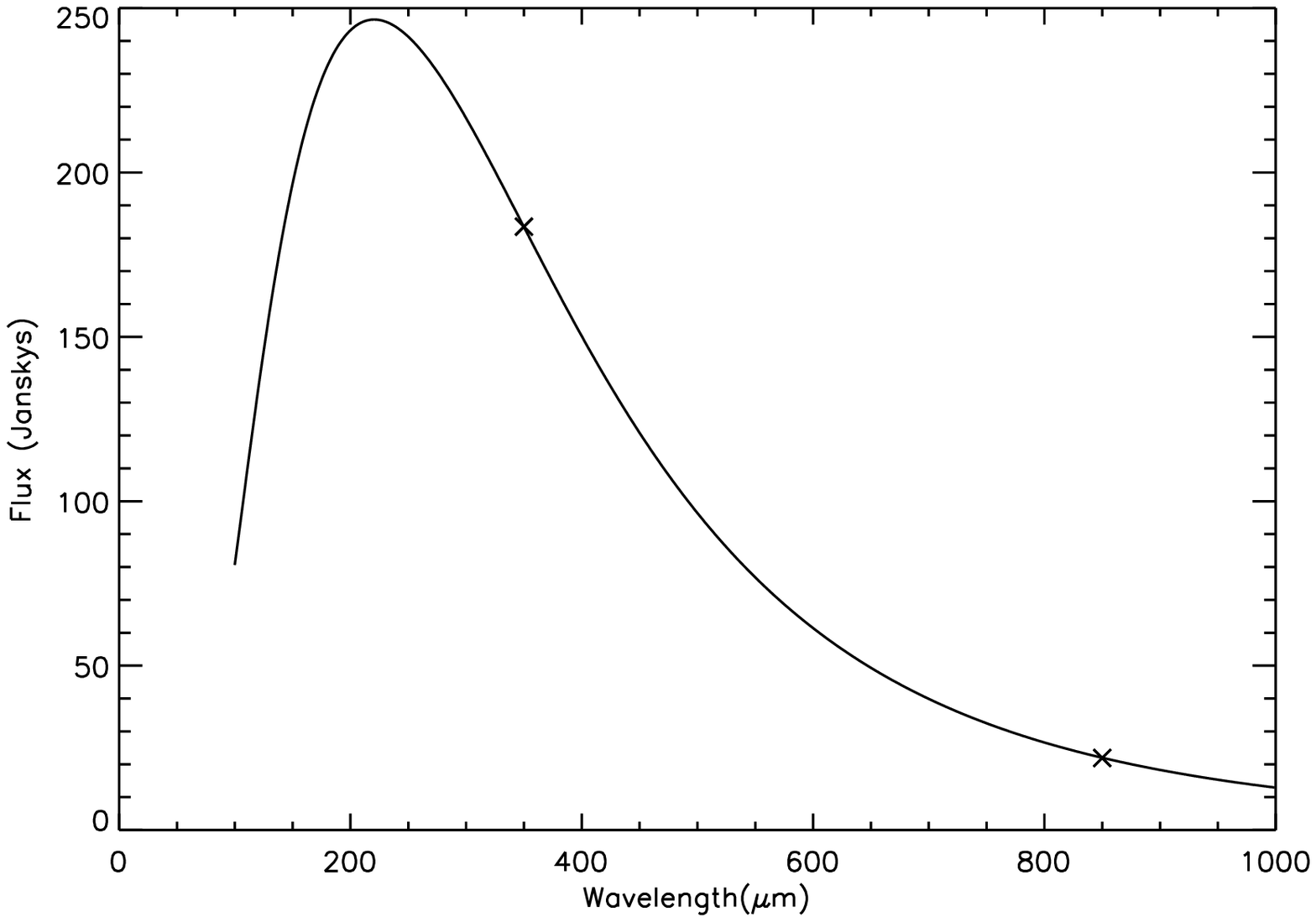}
\caption[Example of Temperature and Optical Depth Fit]{SED of center region of DR21 with fit from Equation \ref{fluxfit}.  x's denote data points and line is fit from Equation \ref{fluxfit}.
\label{ttaufit}}
\end{figure}

The spectral energy distribution at 350 and 850 $\rm\mu m$ was fitted to give estimates of the dust temperature and optical depth at 350 $\rm \mu m$.  The 850 $\rm \mu m$ map was acquired from the SCUBA online archive \footnote{http://www2.cadc-ccda.hia-iha.nrc-cnrc.gc.ca/jcmt} (Project ID: m02bu47).  The SHARC-II map at 350 $\rm \mu m$ was smoothed and repixelated to match the 850 $\rm \mu m$ map.  The intensity at frequency $\rm \nu$ and corresponding wavelength $\rm \lambda$ was then modeled as 

\begin{equation}
I_{\nu}=B_{\nu}(T)(1-e^{-\tau_{350}(350/\lambda)^2}),
\label{fluxfit}
\end{equation}
where the exponent gives the atmospheric transmission at frequency $\nu$ as determined from the transmission at 350 $\rm \mu m$, and $B_\nu(T)$ is the Planck function at frequency $\nu$ and temperature $T$.  An example fit to the data for the center of DR21 is shown in Figure \ref{ttaufit}.  This formula assumes the cloud is optically thin at the wavelengths used.  The ratio of the the 350 $\rm \mu m$ flux to the 850 $\rm \mu m$ is constant throughout the majority of the cloud ($\sim$ 15) with a discrete jump at the edges of the cloud ($\sim$ 25) and a dip in the center ($\sim$ 8).  The constancy of the ratio implies optical thinness except at the center where the ratio drops.   

Temperatures at the peak were $\sim$ 20 K and went up to $\sim$ 25 K in the cloud.  The edges of the cloud where the ratio increased had higher temperature (30-50 K).  The 25 K region of the cloud extended along a rough northeast-southwest direction similar to the direction of the outflow.  The optical depth showed a similar trend with a value at the peak greater than 1.  The optical depth quickly drops below 1 indicating that all but the bright peak areas of DR21 main are optically thin at 350 $\rm \mu m$, as expected from the flux ratios.  
\cite{harvey} found temperatures from 50 $\rm \mu m$ and 100 $\rm \mu m$ flux maps to be $\sim$ 50 K.  These wavelengths may be coming from a different environment with a different temperature.  Supporting a multi-environment hypothesis, \cite{Richardson} found a 350 $\rm \mu m$ to 800 $\rm \mu m$ ratio of $\sim$ 16 at DR21OH with a corresponding temperature of $\sim$ 25 K in regions given to be $\sim$ 37 K by \cite{harvey}.  Throughout the rest of the paper, we used temperatures of 20-25 K from the SHARP and SCUBA fluxes.

\subsection{Mass}
\label{mass}
The mass of cloud within a specified region of a cloud can be determined from its temperature and far infrared flux \citep{hildebrand83}
\begin{equation}
M=\frac{FD^2}{B_{\nu}(T)}C_M
\end{equation}
where F is the flux, D is the distance to the cloud, and
\begin{equation}
C_M=[N(H+H_2)/\tau(\nu)]m_H\mu=1.2\cdot10^{25}(350/400)^{2}m_H\mu,
\end{equation}
where $\tau(\nu)$ is the optical depth of the cloud at frequency $\nu$.  Within a radius of $\sim$ 1 pc (size of SHARP map in E-W direction), DR21 main has a mass of $\sim 25,000 \Msolar$, comparing nicely to the $\sim 20,000 \Msolar$ found by \cite{Richardson}.

\subsection{Magnetic Field Strength}
\label{bfield}
\cite{cf} described a method to estimate the magnetic field in the plane of the sky.  This method related the dispersion of the direction of starlight polarization from a straight line to the strength of the magnetic field,    
\begin{equation}
B_{planeofsky}=\sqrt{\frac{4\pi\rho}{3}}\frac{\sigma_{obs}(v)}{\sigma_{\beta}}
\end{equation} 
where $\rho$ is the density of the gas, $\sigma_{obs}(v)$ is the observed line of sight velocity dispersion, and $\sigma_{\beta}$ is the dispersion in polarization direction in the plane of the sky. The weaker the field, the greater the dispersion of the polarization vector.  

\begin{figure}
\plotone{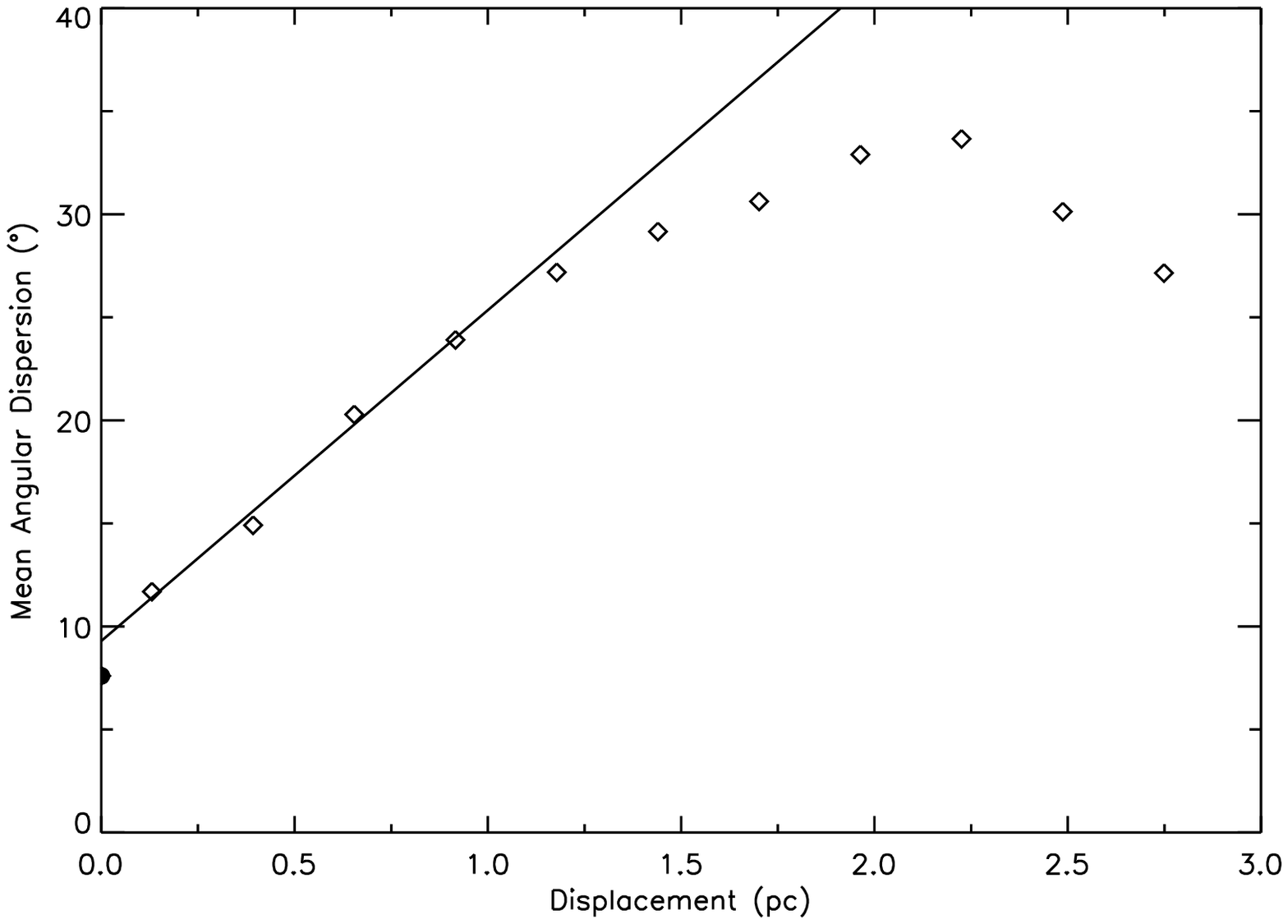}
\plotone{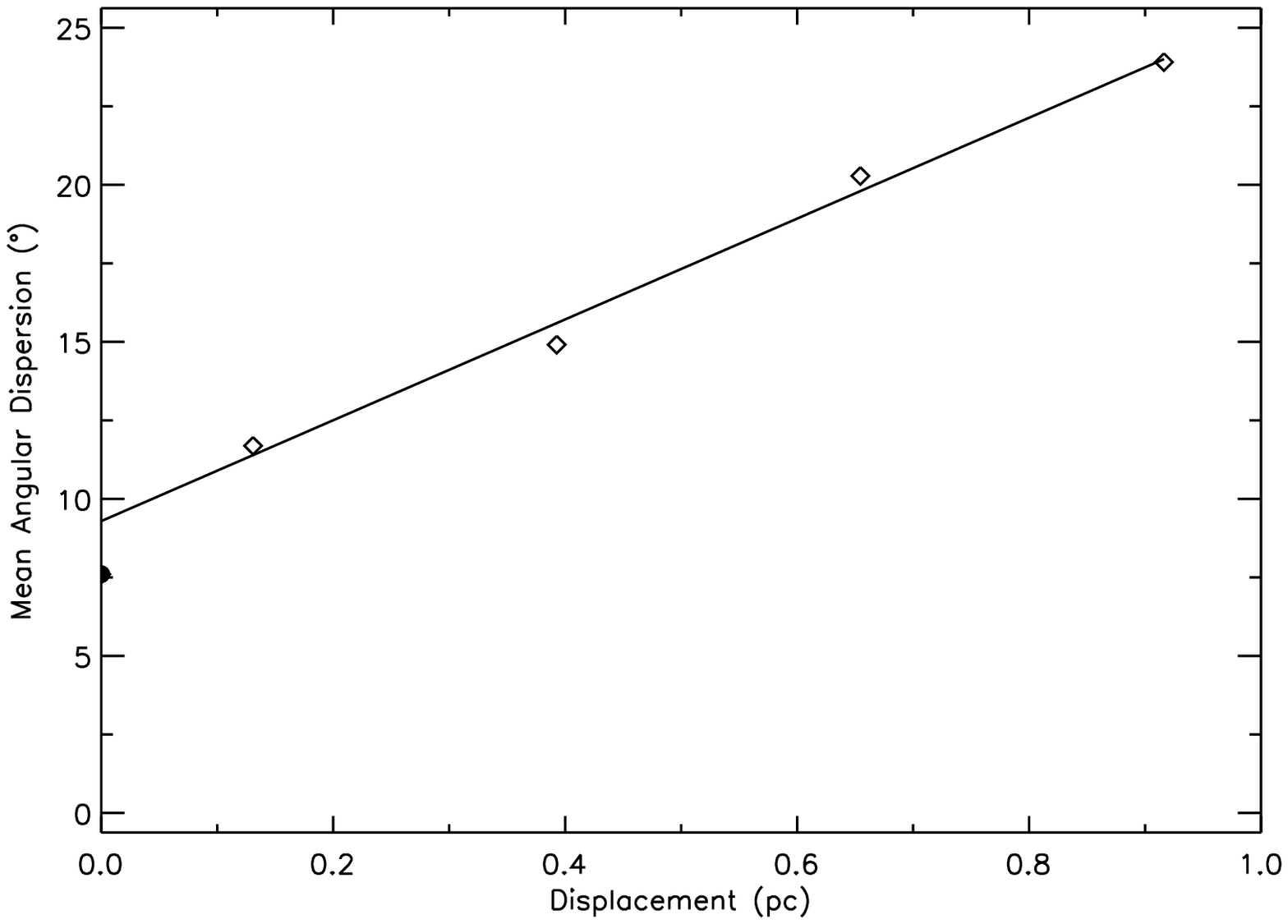}
\caption[Correlation function]{A 2-pt angular correlation function of the polarization angle of DR21 from 2 separate subsets of Hertz data.  The top figure shows the whole cloud and the bottom figure show the same function at separations $\leq \sim \frac{1}{4}$ of the major axis.  The black circle shows value of the function at 0 separation.
\label{hacc}}
\end{figure}

The dispersion of the polarization vectors will be overestimated if it is taken with respect to the mean field.  To determine the dispersion about the local magnetic field one must take into account the inclination of the mean field (\textsection \ref{inclination}) and the structure of the large scale field due to non-turbulent effects such as gravitational collapse, expanding HII regions, and differential rotation (Hildebrand et al. submitted to ApJ).  A polarization map usually shows a smoothly-varying pattern of vectors, therefore at separations small compared to the cloud diameter, the 2-point angular correlation function ($\sqrt<\Delta\beta^2>$) of position angles should increase almost linearly as shown in the bottom panel of Figure \ref{hacc}.  The linear portion would have a zero-intercept if there were no measurement error and no turbulence.  Both effects will cause the whole 2-point correlation to be displaced upward (except at values $<$ correlation length of turbulence which is expected to be much less than the SHARP or Hertz resolution of DR21).  The y-intercept, b, of the best fit line is then the quadratic sum of the estimate of the dispersion, $\sigma_\beta$ and the measurement error, $\sigma_me$.  A cross correlation of repeat observations gives the estimate of the measurement error (value at 0 separation) and an estimate of the quadratic sum of measurement error and turbulent dispersion (intercept of best fit line to linear region at small non-zero separation)   

\begin{equation}
b=\sqrt{\sigma_\beta^2+\sigma_{me}^2}.
\end{equation}

Figure \ref{hacc} shows the correlation function for two subsets of the Hertz data set.  The subsets were constructed by taking every other raw data file starting with file 1 for subset 1 and file 2 for subset 2.  The line best fit to the linear region of the correlation function has an intercept of 9\fdg3.  The value of the correlation function at zero separation is 7\fdg6 giving an estimate of 5\fdg3 for the turbulent dispersion.  Using the Chandrasekhar and Fermi method formula with correction factor from \cite{ost} gives a plane of the sky magnetic field of 3.1 mG where $\sigma_{obs}(v)$ is the observed line of sight velocity dispersion of HCN measured to be $\rm 4.2 kms^{-1}$ and $\rho$ is the mean density calculated from the mass in \textsection \ref{mass} to be $\rm 4.4 x 10^{-19} gcm^{-3}$.  

The method of data acquisition of SHARP prevents repeat observations of the same location (caused by lack of instrument rotator) preventing that sample from being split as done for Hertz.  However, the mean of the individual errors on the angles give a good estimate for the measurement error (7\fdg2 to 7\fdg6 for Hertz).   Using that method of estimating the mean error and fitting a line to the linear region of the correlation function gives a magnetic field estimate of 2.5 mG using SHARP data (dispersion of 6\fdg6).  

Using the measurement of 400 $\rm \mu G$ for the line of sight magnetic field by \citet{Roberts} and taking a $\sim $ 3 mG field in the plane of the sky gives a total magnetic field of $\sim $ 3 mG inclined to the plane of the sky at $\sim $ 10\arcdeg.  This estimate of the angle is probably low due to the Zeeman measurement being from HI, which exists at lower density than the dust we observe in polarization so the field strength is also likely lower there.  Taking the scaling of \citet{HC} and the density of the region observed in HI as $\rm \geq 10^3$, the angle with respect to the plane of the sky is $\rm \leq 53 \arcdeg$.  For a density of $10^4 cm^{-3}$, the angle would be $\sim 20\arcdeg$ \citet{Roberts} assumed that the magnetic field was less than 1 mG because unless the field was very uniform, the field would have bent somewhere such that they detected a 1 mG field.  The dust polarization measurements show a field that is very uniform implying the field could be larger than 1 mG and not detected in the Zeeman observations.

\subsection{3-Dimensional Model}
\label{3d}
\begin{figure}
\plotone{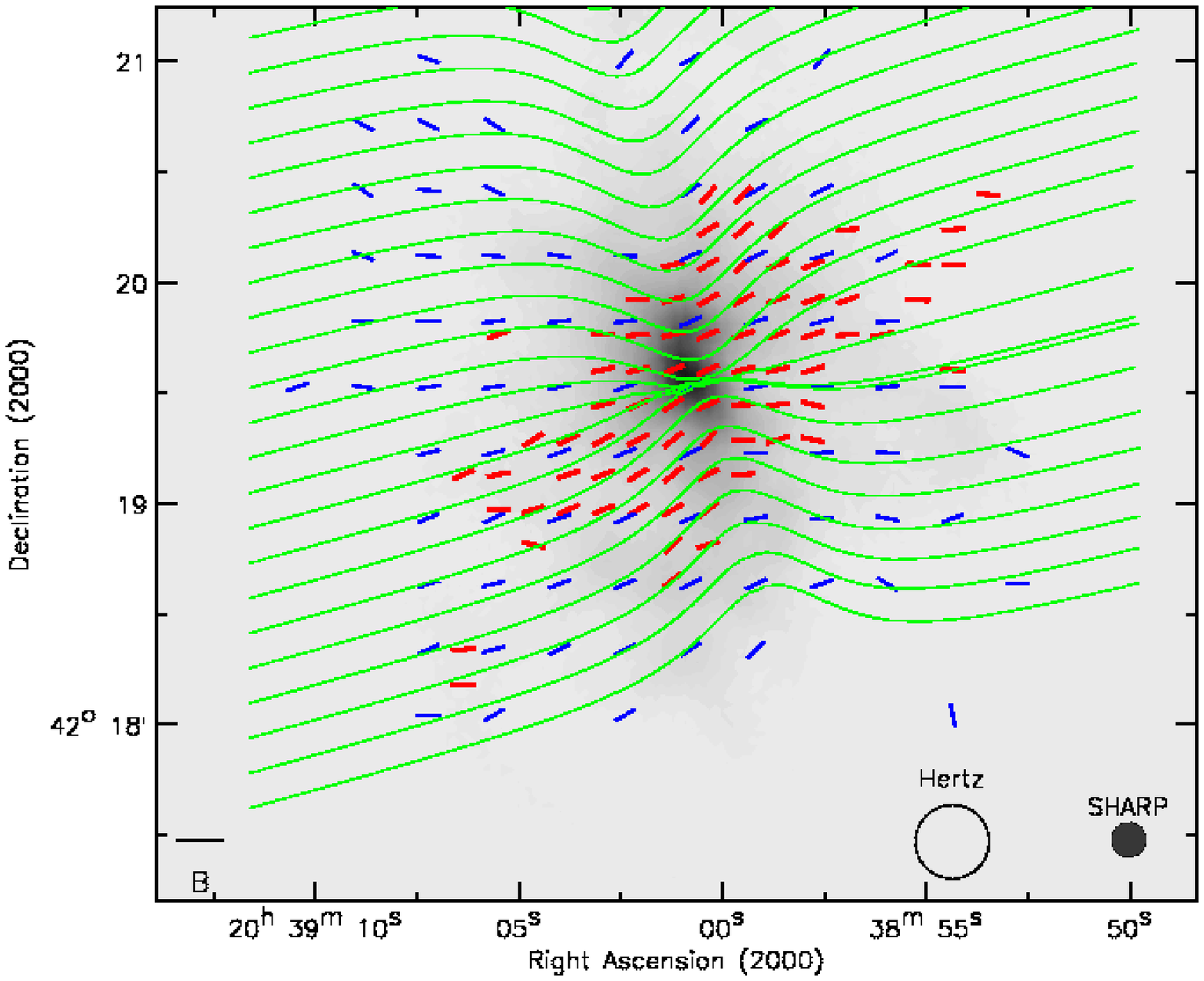}
\caption[3D fit]{350 $\rm \mu m$ SHARP polarization map (red vectors) with overplot of projected plane of sky vectors from 3D fit.
\label{3dfit}}
\end{figure}

A three dimensional model of the flux and magnetic field was constructed and fit to the observed flux and polarization data.  The flux was modeled as a Gaussian ellipsis of the form
\begin{equation}
F\propto exp[-\frac{x^2}{a^2} - \frac{y^2+z^2}{b^2}],
\end{equation}
where $x$ is coordinate along the symmetry axis of the magnetic field, $y$ and $z$ are the other two Cartesian coordinates, and $a$ and $b$ are widths of the Gaussians to be fit.  The magnetic field was modeled as 
\begin{equation}
B = \left( B_x,\frac{d}{dx}(\frac{cx^2}{x^2+d})\frac{y}{\sqrt{y^2+z^2}},\frac{d}{dx}(\frac{cx^2}{x^2+d})\frac{z}{\sqrt{y^2+z^2}} \right),
\end{equation}
where 
\begin{equation}
B_x \propto \int_{-\infty}^{\infty}F dx.
\end{equation} 
This form was chosen to give a magnetic field with the shape of an hourglass that straightened at large distance from the center.  The magnetic field strength in the x direction in the midplane was set proportional to the column density.  Assuming flux freezing, the magnetic field strength in the x direction for locations outside the midplane was set to a value in the midplane that is found by tracing the field line back to the midplane.  The field in the y-z plane was assumed radial and the strength was found from the strength in the x direction and the overall magnetic field direction.  The flux and magnetic field strengths were then rotated and summed along the line of sight.  The values of $a$, $b$, $c$, and $d$ were then set to give the best fit to the flux and the magnetic field direction determined by the polarization measurements.  The best fit values were $a=5, b=15, c=10, d=90$ (units for $a,b,c,d$ are SHARP pixels = 2\farcs375, 20\arcdeg for the angle of the magnetic field with the plane of the sky, and 15\arcdeg East of North for the angle of the projected minor axis on the plane of the sky.  The fit for the magnetic field direction was quite good (chi-squared of $\sim$ 2).  The flux fit was poor as DR21-Main is not an exact ellipse.  Figure \ref{3dfit} shows the model vectors along with the SHARP vectors.  The angle with the plane of the sky is somewhat higher than found in \textsection \ref{bfield} but agrees well with the measured inclination angles farthest from the center (see \textsection \ref{inclination}).  

\subsection{Gravitational and Magnetic Potential Energies}
\begin{figure}
\plotone{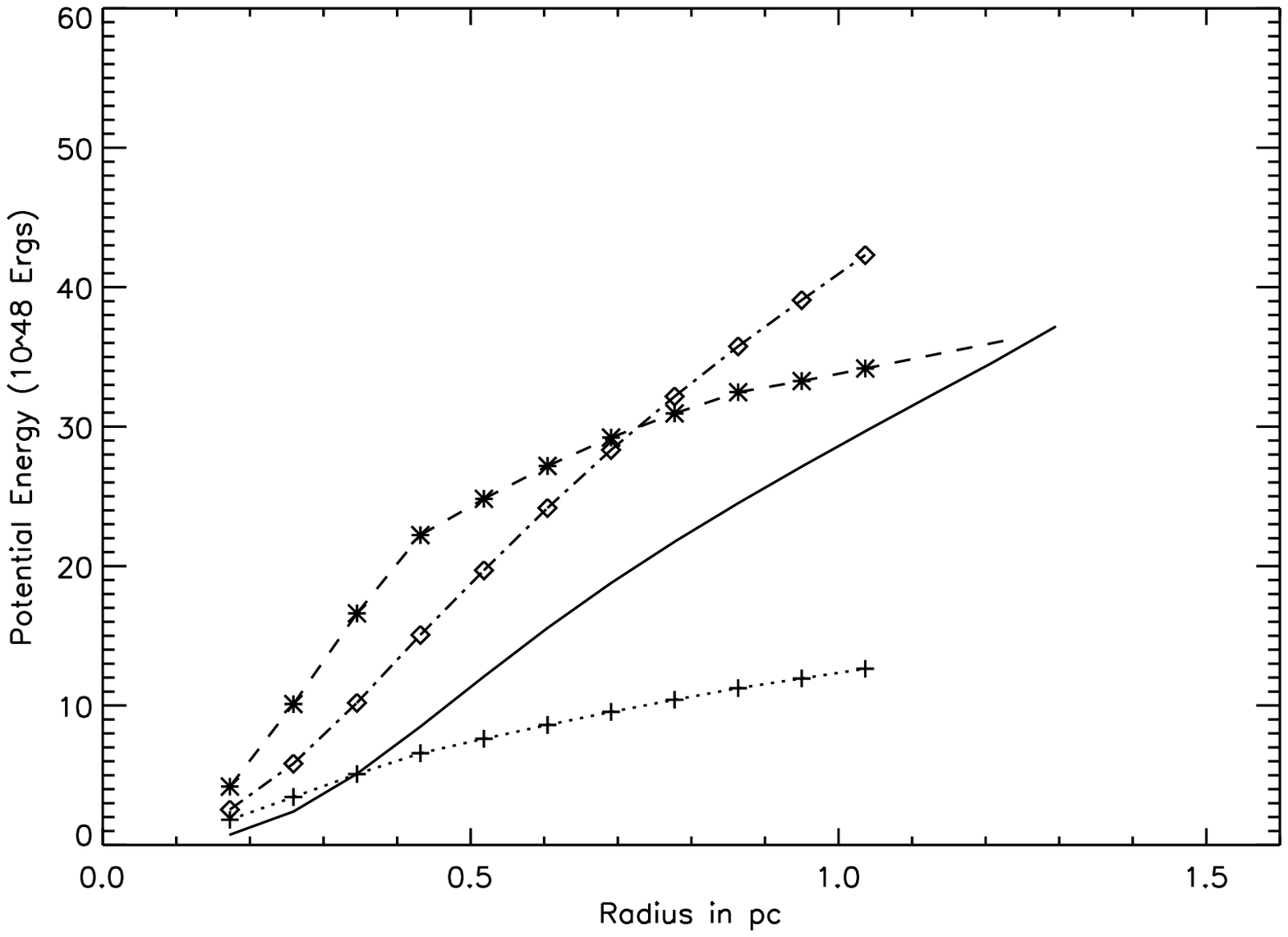}
\caption[Potential Energy versus radius for DR21] {The gravitational potential (dashed line and asterisks) is computed from the calculation of mass (\textsection \ref{mass}).  The magnetic field energy (solid line) is based on determination of magnetic field strength using 2-dimensional CF method and published Zeeman measurements (\textsection \ref{bfield}) assuming equipartion between the uniform and fluctuating field \citep{novak04}.  Recent results (Hildbrand et al. in preparation) show that equipartition is an overestimate and that the ratio of the turbulent to mean field is approximately 10\%.  The turbulent energy (dotted line and crosses) is calculated from the mass and observed HCN line widths.  The dot-dashed line and squares is the sum of magnetic and turbulent energies.
\label{gvsbimage}}
\end{figure}

\begin{figure}
\plotone{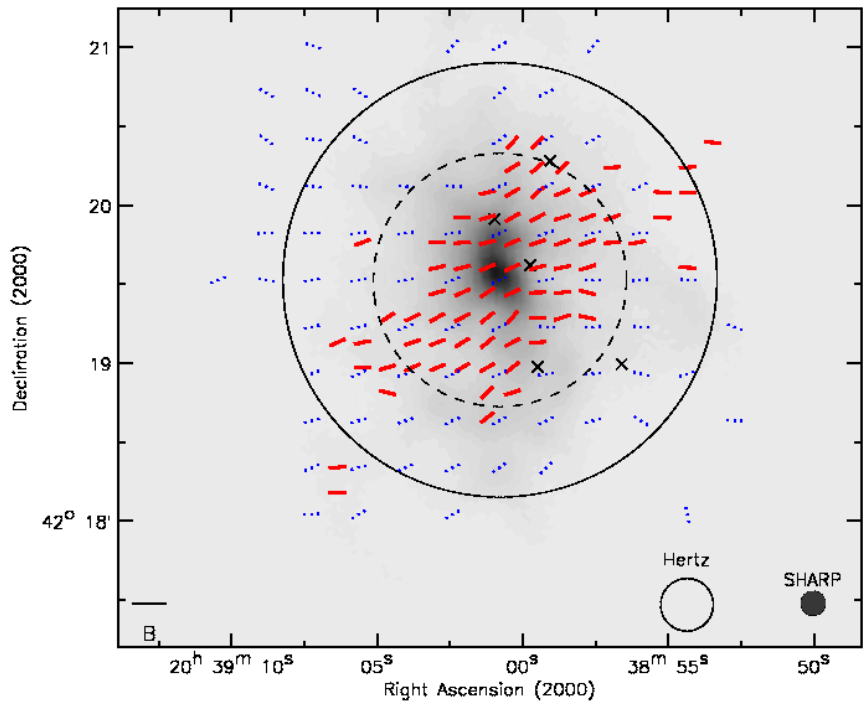}
\caption[Magnetic field vectors and energy-equivalent radii] {DR21M with SHARP magnetic field vectors.  Solid circle delineates radius where magnetic energy becomes larger than gravitational potential energy.  Dotted circle delineates radius where the sum of magnetic and turbulent energy becomes larger than gravitational potential energy assuming equipartion.  The grayscale background is from SHARC-II and black 'x''s mark the location of Spitzer point objects.
\label{dr21_bb}}
\end{figure}

\begin{figure} 
\plotone{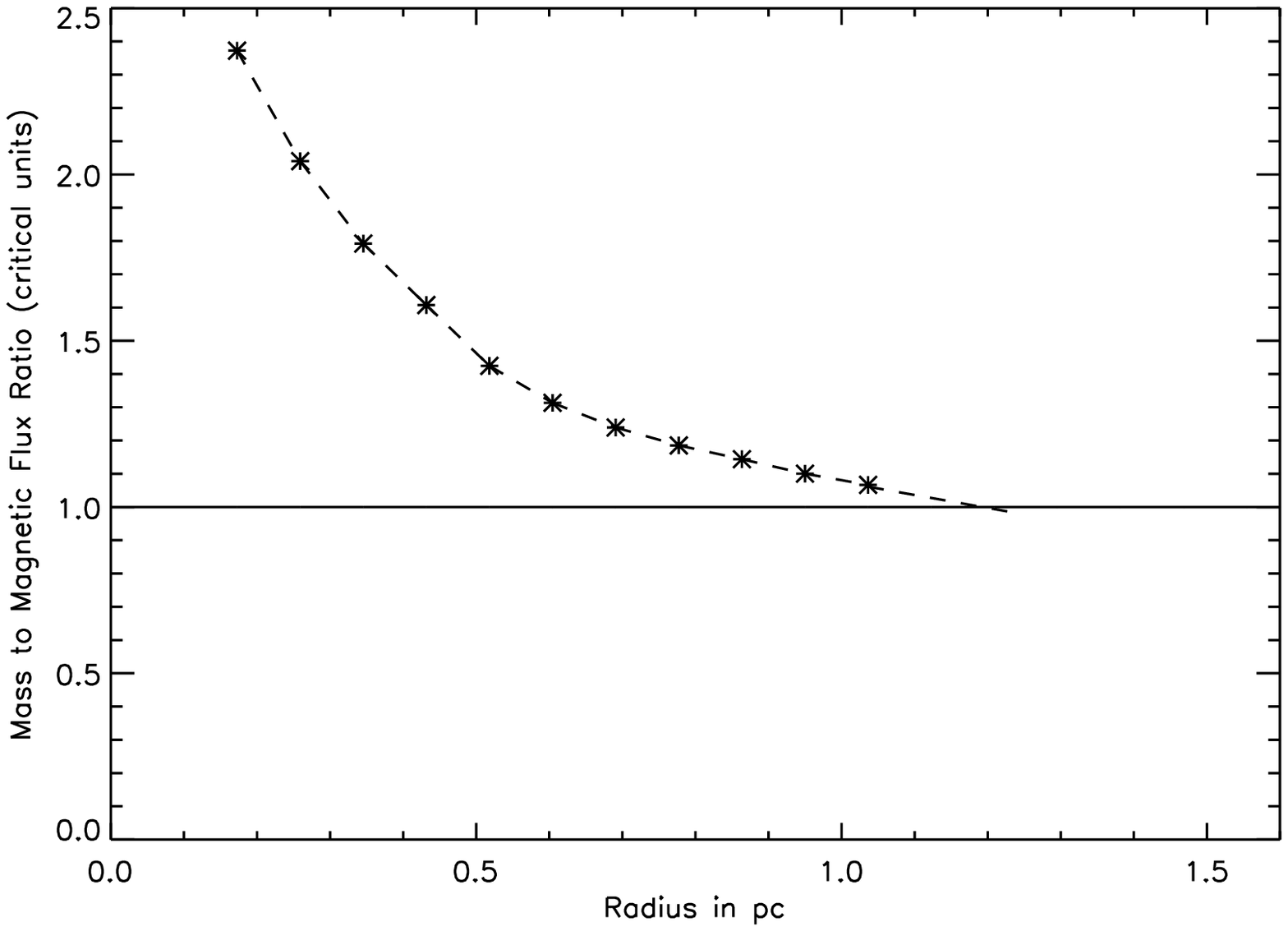} 
\caption[Mass to Magnetic Flux ratio of DR21M] {Mass to flux ratio in units of the critical ratio between supercritical and subcritical.  Asterisks are calculations of ratio and dashed line is interpolation/extrapolation  
\label{mtobf}} 
\end{figure}

Using the mass and magnetic field estimates discussed in \textsection \ref{mass}, \textsection \ref{bfield}, and the 3D model presented in \textsection \ref{3d}, one can estimate the dependence on radius of the gravitational and magnetic potential energies.  The gravitational potential energy of a flattened centrally condensed spheroid at a radius R centered on the peak of the cloud is
\begin{equation}
E_{grav}=\frac{3aGM^2}{5R},
\end{equation}
and the magnetic energy of the uniform field is given by 
\begin{equation}
E_{mag}=\frac{B^2R^3b}{6},
\end{equation}
where the values $a=1.2$ and $b=0.3$ and are calculated by setting the energies equal when the mass to magnetic flux ratio is the critical value \citep{mckee}.  B is scaled according the model in \textsection \ref{3d} and normalized such that the energy inside 1 pc is the same as for the mean field calculated in \textsection \ref{bfield}.  By comparing large scale polarization maps to with magnetohydrodynamic turbulence simulations, \citet{novak04} found that the ratios of the uniform to fluctuating magnetic field in the clouds NGC 6334 and G333.6-0.2  were in the range 0.6-2.0.  Figure \ref{gvsbimage} shows the gravitational and magnetic potential energies (assuming a ratio between uniform to fluctuating field of 1) versus radius.  Extrapolating the gravitational potential energy curve to find the radius at which the two energies are equal gives an estimate of $\sim$ 1.2 parsec.   The dotted line in Figure \ref{gvsbimage} shows turbulent energy calculated from the average HCN line widths of 4.2 km/s and assuming the velocities are isotropic.  The sum of these two energies is the dotted-dashed line and intersects the gravitational energy curve at 0.7 parsecs.  One would then expect the critical radius to be between 0.7 and 1.2 parsecs.  The critical radius should not be regarded as the radius which is supported against collapse since the virial theorem can not be applied in this way to magnetic clouds \citep{Mous2,Mous1,dib}.  Figure \ref{mtobf} shows a plot of the mass to magnetic flux ratio versus radius.  The ratio is in units of the critical value of the ratio defined by \citet{Mouandspit} 
\begin{equation}
M/\Phi_B=c_{\Phi}/\sqrt{G},
\end{equation}
where $c_{\Phi}$ is taken to be 0.12 \citep{Tomisaka}.  The ratio dips below the critical ratio at $\sim$ 1.2 parsec.  The mass inside this region is $\rm \sim 25,000 \Msolar$ (\textsection \ref{mass}) which is approximately the entire mass of the cloud.  Figure \ref{dr21_bb} shows the polarization vectors with circles at the critical radii with and without turbulent energy.  The magnetic field configuration changes at around these radii from the pinched structure to the east-west mean direction (see Figure \ref{sharpandhertz-fits}).  This geometry implies that the cloud is relatively stable against collapse at these radii, implying that the magnetic field is providing a significant amount of support against collapse. However the presence of protostars are radii greater than 1.2 parsec in Figure \ref{dr21_bb} suggests that the true picture may be more complex.


\section{Summary}
We present a three-dimensional model of the molecular cloud DR21 (Main).  Polarimetry at 350 $\rm \mu m$ has provided a map of the magnetic field as projected on the plane of the sky; line observations of ion and neutral molecules have provided measurements of the inclination of the field to the line of sight; photometry has provided column densities and $\rm 350\mu m /850 \mu m$ color temperatures.

A 3-dimensional model of an hourglass configuration magnetic field inclined to the line of sight is fit to the observations.  We find a pinched field with an axis of 10\arcdeg-20\arcdeg \ from the plane of the sky.  The mean field strength is $\sim$ 3 mG.  The gravitational potential energy is equivalent to energies from support mechanisms (magnetic field and turbulence) at a radius between $\rm \sim 0.8 pc - 1.1 pc$, a region encompassing $\rm \sim 20,000 \Msolar$.

\acknowledgments{I am grateful to my adviser Roger Hildebrand and our collaborators Michael Attard, David Chuss, Jackie Davidson, Jessie Dotson, Darren Dowell, Martin Houde, Megan Krejny, Lerothodi Leeuw, Hua-bai Li, Giles Novak, Hiroko Shinnaga, Konstantinos Tassis and John Vaillancourt for support, observing assistance, and critical comments on drafts of this paper.  Observations at the Caltech Submillimeter Observatory would not have been possible without the assistance of the CSO staff.  I would also like to thank S. S. Meyer, A Konigl, and S. P. Swordy for guidance and critical suggestions in the preparation of this paper.   The author has been supported by NSF grant AST 05-05230 and the University of Chicago.  Work with the SHARP polarimeter has been funded by AST 02-41356, 05-05230, 05-05124.  The CSO is funded by the NSF through grant AST 05-40882}

\end{document}